\RequirePackage{silence}
\documentclass[fleqn,usenatbib,useAMS]{mnras} 
\usepackage{graphicx}
\usepackage{amsmath}

\usepackage{amsfonts}
\usepackage{float}
\usepackage{bm}
\usepackage{cancel}
\usepackage[perpage,symbol*]{footmisc}
\usepackage{ae,aecompl}
\usepackage{array}
\usepackage{soul}
\usepackage{mathtools}
\usepackage{multirow}
\usepackage[utf8]{inputenc}
\usepackage{booktabs}
\usepackage{graphicx}
\usepackage{subcaption}
\usepackage{makecell}

\DeclareRobustCommand*{\matr}[1]{\mathbfss{#1}}

\title[Overestimated inclinations of disc galaxies]{Overestimated inclinations of Milgromian disc galaxies: the case of the ultradiffuse galaxy AGC 114905} 

\author[I. Banik et al.]{Indranil Banik$^1$\thanks{Email: \href{mailto:ib45@st-andrews.ac.uk}{ib45@st-andrews.ac.uk} (Indranil Banik),\newline $~~~~~~~~$ \href{mailto:tnsrikanth1998@gmail.com}{tnsrikanth1998@gmail.com} (Srikanth T. Nagesh)}, Srikanth. T. Nagesh$^{2,3}$, Hosein Haghi$^{3,4}$, Pavel Kroupa$^{3,5}$
\newauthor
and Hongsheng Zhao$^1$ \vspace{10pt} \\
$^{1}$Scottish Universities Physics Alliance, University of Saint Andrews, North Haugh, Saint Andrews, Fife, KY16 9SS, UK \\
$^{2}$Argelander-Institut f\"ur Astronomie, Universit\"at Bonn, Auf dem H\"ugel 71, D-53121 Bonn, Germany \\
$^{3}$Helmholtz-Institut f\"ur Strahlen und Kernphysik (HISKP), University of Bonn, Nussallee 14$-$16, D-53115 Bonn, Germany \\
$^{4}$Department of Physics, Institute for Advanced Studies in Basic Sciences (IASBS), PO Box 11365-9161, Zanjan, Iran \\
$^{5}$Astronomical Institute, Faculty of Mathematics and Physics, Charles University, V Hole\v{s}ovi\v{c}k\'ach 2, CZ-180 00 Praha 8, Czech Republic}

\pubyear{2022}
\begin{document}
\label{firstpage}
\pagerange{\pageref{firstpage}--\pageref{lastpage}}

\maketitle

\begin{abstract} 
We present two hydrodynamical star-forming simulations in the Milgromian dynamics (MOND) framework of a gas-rich disc galaxy with properties similar to AGC 114905, which has recently been argued to have a rotation curve (RC) that is inconsistent with the MOND prediction. Our first model considers the galaxy in isolation, while our second model includes an external field of $0.05 \, a_{_0}$, the estimated gravitational field from large-scale structure. We show that isophotes in the face-on view can differ from circular at the 50\% level. This could mislead observers into overestimating the inclination $i$ between disc and sky planes. Because RCs require a correction factor of $1/\sin i$, the actual RC could be much higher than reported by observers. This plausibly reconciles AGC 114905 with MOND expectations.

\end{abstract}


\section{Introduction}
\label{Introduction}

The observed rotation curves (RCs) of galaxies are flat in their outskirts \citep[e.g.][and references therein]{Faber_1979}. This strongly contradicts the expected Keplerian decline beyond the extent of the luminous matter, where the Newtonian inverse square law implies that the circular rotation velocity $v_c$ scales with galactocentric radius $r$ as $v_c \propto 1/\sqrt{r}$. This acceleration discrepancy could indicate that galaxies are embedded in massive extended haloes of cold dark matter \citep[CDM;][]{Ostriker_Peebles_1973}. These haloes are nowadays a cornerstone of the standard cosmological framework known as $\Lambda$CDM \citep*{Efstathiou_1990, Ostriker_Steinhardt_1995}. However, several very serious problems have been identified with $\Lambda$CDM on galactic scales \citep{Peebles_2010, Kroupa_2010, Kroupa_2012, Kroupa_2015, Bullock_2017, Bose_2019, Haslbauer_2019_DF2, Pawlowski_2021}. It also faces serious problems on cosmological scales, including the Hubble tension \citep{Valentino_2021} but also with regard to voids \citep*{Haslbauer_2020} and galaxy clusters \citep*{Asencio_2021} being more extreme than predicted in this framework.

$\Lambda$CDM relies on the standard law of gravitation, which was formulated using only Solar System data. Rather than postulating the existence of exotic CDM particles beyond the well-tested standard model of particle physics, \citet{Milgrom_1983} noted that the acceleration discrepancies arise when the gravitational field $g$ is much weaker than in the Solar System. He introduced a new constant of nature ($a_{_0}$) with dimensions of an acceleration such that in an isolated spherically symmetric system with Newtonian gravity $g_{_N}$,
\begin{eqnarray}
	g ~=~ \left\{
	\begin{array}{ll}
		\sqrt{g_{_N} a_{_0}} &\quad \left( g_{_N} \ll a_{_0} \right) \, , \\ 
		g_{_N} &\quad \left( g_{_N} \gg a_{_0} \right) \, .
	\end{array}
	\right.
	\label{MOND_basic}
\end{eqnarray}
This Milgromian dynamics (MOND) theory is one of the most competitive alternative solutions to the acceleration discrepancy problem, as extensively reviewed in \citet{Famaey_McGaugh_2012}. Empirically, $a_{_0} = 1.2 \times 10^{-10}$~m/s\textsuperscript{2} or 3.9~pc/Myr\textsuperscript{2} \citep*{Begeman_1991}, with more recent studies confirming this value \citep*{Gentile_2011, McGaugh_Lelli_2016}. In the deep-MOND limit, MOND satisfies a principle known as spacetime scale invariance according to which any trajectory $\left( t, \bm{r} \right)$ that solves the equations of motion implies that these are also solved by another trajectory $\left( \lambda t, \lambda \bm{r} \right)$ for any constant $\lambda$ \citep{Milgrom_2009_DML}. Considering the case of uniform circular motion reveals that the RC of an isolated point mass must be flat in the deep-MOND regime. Newtonian dynamics does not obey spacetime scale invariance.

The limits in Equation~\ref{MOND_basic} follow from a new non-relativistic gravity theory based on a Lagrangian, thereby respecting the usual conservation laws with respect to linear and angular momentum and the energy \citep{Bekenstein_Milgrom_1984}. The generalized non-linear Poisson equation derived from this Lagrangian leads to a rich behaviour of gravitating systems, accounting automatically for hitherto unexplained phenomena like the KBC void and Hubble tension \citep{Haslbauer_2020}, extreme galaxy cluster collisions at high redshift \citep{Katz_2013, Asencio_2021}, the Local Group satellite planes \citep{Banik_Ryan_2018, Bilek_2018, Bilek_2021, Banik_2022_satellite_plane}, and many others besides while remaining consistent with galaxy cluster dynamics and early Universe observations \citep[for an extensive review, see][]{Banik_Zhao_2022}. The MOND Poisson equation can be solved using standard mesh-refinement codes like \textsc{phantom of ramses} \citep[\textsc{por};][]{Lughausen_2015, Nagesh_2021} and \textsc{raymond} \citep{Candlish_2015}.

A robust prediction of the MOND Poisson equation is that galaxy RCs become asymptotically flat at some velocity $v_{_f}$ which depends on the baryonic mass $M_b$ according to
\begin{eqnarray}
    v_{_f} ~=~ \left( G M_b a_{_0} \right)^{1/4} \, .
    \label{asymptoticvelocity}
\end{eqnarray}
More generally, MOND predicts a very tight radial acceleration relation (RAR) between $g$ as deduced from RCs and $g_{_N}$ as deduced from the distribution of observable matter. A tight RAR is indeed apparent observationally \citep{McGaugh_Lelli_2016, Lelli_2017} from the Spitzer Photometry and Accurate Rotation Curves (SPARC) dataset \citep*{SPARC}. RCs offer strong tests of the MOND paradigm, especially if the central surface density $\Sigma_0 \ll \Sigma_{M}$, where the critical surface density in MOND is
\begin{eqnarray}
    \Sigma_M ~\equiv~ \frac{a_{_0}}{2 \mathrm{\pi} G} ~=~137 \, M_\odot/\text{pc}^2 \, .
    \label{Sigma_M}
\end{eqnarray}
This is because $g \gg g{_N}$ in such low surface brightness (LSB) galaxies, making them well suited for testing different explanations for the galactic acceleration discrepancies \citep{McGaugh_2020}.

In a recent paper, \citet{Mancera_2022} report new HI interferometric observations of the gas-rich ultra-diffuse galaxy AGC 114905, tracing its HI emission up to 10 kpc from the galaxy centre. They infer from these a surprisingly low $v_{_f} \approx 23$~km/s. They then deduce that the observed RC of AGC 114905 can be explained almost entirely by $g_{_N}$ of the baryons alone, with little room for CDM within the outermost observed radius. They find that the RC can only be reproduced by standard CDM haloes if the halo concentration is as low as 0.3, completely off $\Lambda$CDM expectations. They also argue that the observed RC significantly disagrees with the normalization and shape of the MOND RC.

In this paper, we use fully self-consistent hydrodynamical MOND simulations that include star formation and supernova feedback to show that LSB Milgromian disc galaxies with properties tailored to those of AGC 114905 are often not round when viewed face-on. This can lead to the inclination $i$ between disc and sky planes appearing to be much higher than it actually is. This is critical because the claim of \citet{Mancera_2022} to have falsified MOND relies on the correctness of their measured $i = 32^\circ$, but those authors showed that consistency is gained for an almost face-on $i = 11^\circ$. We argue that this is quite possible, so the kinematics of AGC 114905 are consistent with MOND. Moreover, even a small amount of warping in such a nearly face-on galaxy could seriously impact the shape of the RC deduced from line-of-sight velocities.

This paper is structured as follows: In Section~\ref{Methods}, we describe the simulations and how we analyse them. We then present our results in Section~\ref{Results} and discuss their implications in Section~\ref{Discussion}. Our conclusions are given in Section~\ref{Conclusions}. Videos of our simulations are publicly available.\footnote{
Isolated face-on: \url{https://youtu.be/VkZq2Se0_lU}\\
With an external field, face-on: \url{https://youtu.be/3wC3yRkrXnc}}

\section{Methods}
\label{Methods}


\subsection{MOND galaxy simulations}
\label{POR_simulations}

\begin{table}
    \begin{tabular}{ll}
    \hline
    Parameter & Value \\ \hline
    Total mass & $1.42 \times 10^9 M_\odot$ \\
    Stellar mass & $1.30 \times 10^8 M_\odot$ \\
    Gas fraction & 0.90 \\
    Stellar disc scale length & 1.79 kpc \\
    Gas disc scale length & 4.47 kpc \\
    $\Sigma_0/\Sigma_M$ & 0.122 \\
    $v_{_f}$ & 68.96 km/s \\
    Best resolution & 195 pc \\ \hline
    \end{tabular}
    \caption{Parameters of our galaxy models. The isolated model assumes $g_{\textrm{ext}} = 0$, while the EFE model assumes $g_{\textrm{ext}} = 0.05 \, a_{_0}$ directed at $30^\circ$ to the disc plane but with no component along $y$.}
    \label{parameters}
\end{table}

The galaxy models presented here were evolved for 5~Gyr. They have two main components: an inner stellar disc with scale length $r_d$ and an outer gas disc with scale length $r_g = 2.5 \, r_d$ (this factor will be justified in a future work; see S. T. Nagesh et al. 2022, in preparation). Both discs have an exponential surface density profile initially \citep{Freeman_1970}. Their adopted parameters (listed in Table \ref{parameters}) were chosen based on the properties of AGC 114905 discussed in section 2 of \citet{Mancera_2022}. The models are entirely in the MOND regime as the central surface density $\Sigma_0$ is below the MOND surface density $\Sigma_{M}$ (Equation~\ref{Sigma_M}): the ratio $\Sigma_{0}/\Sigma_{M} = 0.12$ initially (see Table \ref{parameters}). For reference, this ratio was 10 in the MOND model presented in \citet{Roshan_2021_disc_stability}, while the M33 model presented in \citet{Banik_2020_M33} used a lower ratio of $\approx 2$ (see their figure~1). Thus, our work considers a much lower $\Sigma_{0}$ than these earlier MOND simulations, which is needed to match the properties of AGC 114905. Note that since the equations of motion become self-similar in the deep-MOND regime due to the spacetime scale invariance discussed in Section~\ref{Introduction}, our isolated results are also applicable to galaxies with a much lower $\Sigma_0$. However, in the presence of an external field, its strength relative to the internal gravity of the disc provides another dimensionless parameter (Section~\ref{Including_EFE}).

The axisymmetric initial conditions for these galaxy models were generated using Disk Initial Conditions Environment \citep[\textsc{dice};][]{Perret_2014}. \textsc{dice} was modified to initialize discs in Milgromian gravity \citep[\textsc{m-dice};][]{Banik_2020_M33}. \textsc{m-dice} is publicly available.\footnote{\label{PoRbit}\url{https://bitbucket.org/SrikanthTN/bonnPoR/src/master/}} We use the hydrodynamic version of this \citep[\textsc{h-dice};][]{Nagesh_2021}, with the gas temperature settings being the same as in \citet{Banik_2020_M33}.

These disc galaxy templates were advanced using \textsc{por}, which numerically implements the quasi-linear formulation of MOND \citep[QUMOND;][]{QUMOND} by a patch to the 2015 version of the hydrodynamics and $N$-body solver \textsc{ramses} \citep{Teyssier_2002}. The governing equation of QUMOND is:
\begin{eqnarray}
    \nabla \cdot \bm{g} ~=~ \nabla \cdot \left( \nu \bm{g}_{_N} \right) \, , \quad \nu ~=~ \frac{1}{2} + \sqrt{\frac{1}{4} + \frac{a_{_0}}{g_{_N}}} \, .
    \label{Simple_nu}
\end{eqnarray}
We transition between the asymptotic limits in Equation~\ref{MOND_basic} using the simple interpolating function as this works well observationally \citep{Famaey_Binney_2005, Iocco_Bertone_2015, Lelli_2017, Banik_2018_Centauri}. Notice that $\bm{g}_{_N}$ first has to be determined from the baryonic distribution using standard techniques, allowing the algorithm to find $\nu$ at every point on the grid and thereby algebraically determine the source term for $\bm{g}$. This is found subject to the boundary condition that the potential $\Phi$ is asymptotically given by:
\begin{eqnarray}
    \Phi ~=~ \sqrt{GMa_{_0}} \ln R \, ,
    \label{Phi_iso}
\end{eqnarray}
where $M$ is the total mass inside the simulation volume and $R$ is the distance from the barycentre in the simulation unit of length, the choice of which has no bearing on the results. Only a standard Poisson solver is required, greatly reducing the computational difficulties compared to the earlier version of MOND \citep{Bekenstein_Milgrom_1984}.

\textsc{por} is publicly available$^{\ref{PoRbit}}$ along with a user manual to set up isolated and interacting disc galaxy simulations and extract results from them into human-readable form \citep{Nagesh_2021}. We extract the particle and gas data from the outputs at 100~Myr intervals using \textsc{extract\_por} and \textsc{rdramses}, respectively. Both codes are publicly available.$^{\ref{PoRbit}}$

We consider the face-on view of each galaxy for a distant observer along the $z$-axis. To do this, we first find the combined barycentre of the stars and gas. We subtract its position and velocity from the gas cell data and then bin the gas cells according to their central $\left( x, y \right)$. In this way, we build up a two-dimensional (2D) distribution of the gas surface density $\Sigma_{\text{gas}}$. We then plot contours on this image and analyse a representative contour in the outer region, as described next. The level of this outer contour is chosen so its size is similar to the outermost isophote used by \citet{Mancera_2022} to estimate $i$ (see their figure~7).

\subsubsection{Including the external field effect (EFE)}
\label{Including_EFE}

The non-linearity of MOND gives rise to the so-called EFE, which causes the self-gravity of a system to be weakened if the system as a whole is accelerated uniformly by an external field $\bm{g}_{\textrm{ext}}$, even in the complete absence of tidal effects. This can be understood using Equation~\ref{Simple_nu}: the argument of $\nu$ includes the Newtonian gravity due to both the system and external sources beyond it. The EFE is a fundamental consequence of MOND that has always been part and parcel of it \citep[see section~3 of][]{Milgrom_1983}. Observational evidence for the EFE is discussed further in section~3.3 of \citet{Banik_Zhao_2022}. Disc galaxies experiencing the strong EFE of a cluster environment can become lopsided \citep{Candlish_2018}, though even a rather weak EFE can have subtle consequences \citep{Banik_2020_M33}.

We include the EFE in our simulations using the technique discussed in \citet{Banik_2020_M33}. As in that work, we assume $\bm{g}_{_\textrm{ext}}$ lies entirely within the $xz$ plane and is directed $30^\circ$ above the disc plane. We take $g_{_\textrm{ext}} = 0.05 \, a_{_0}$ since this is roughly the preferred value for the EFE on very large scales in a MOND model of the local supervoid which solves the Hubble tension \citep{Haslbauer_2020}. Thus, even a very isolated galaxy should experience an external field of a few percent of $a_{_0}$. If the galaxy has a very low surface density such that $\Sigma_0/\Sigma_M \ll 0.05$, then the EFE would dominate everywhere, making the galaxy quasi-Newtonian \citep[see section 2.4 of][]{Banik_Zhao_2022}. Since the equations of motion in this regime are like those of Newtonian dynamics but with renormalized $G$, the disc should be unstable for much the same reason that a bare disc is unstable in Newtonian gravity \citep{Hohl_1971}. Thus, MOND could struggle to explain a thin disc with a sufficiently low surface brightness, though this would depend on the estimated $g_{\textrm{ext}}$.

In our EFE model, the barycentre drifts due to numerical effects \citep[see footnote 14 of][]{Banik_2020_M33}. We found that the position of the barycentre changes very nearly quadratically with time, indicating an acceleration of $1.6 \times 10^{-3} \, a_{_0}$ in a direction almost opposite to that of $\bm{g}_{\textrm{ext}}$ and thus almost entirely within the $xz$ plane. This is expected in a problem that is symmetric with respect to $\pm y$. The barycentre drift can be reduced with a larger box, but the resolution would then be poorer. The only way around this problem is to raise the maximum number of refinement levels, but this is quite computationally intensive. Another issue is that the disc precesses by $\approx 11^\circ$ by the end of the simulation, so we rotated the coordinate system in our EFE model such that the images are the view for an observer located very far away in the direction of the total angular momentum.

\subsection{Fitting a contour with a circle and an ellipse}
\label{Circle_ellipse_fitting}

We fit a circle and an ellipse to a representative contour in $\Sigma_{\text{gas}}$. For this, we analyse the list of $\left( x_i, y_i \right)$, where the integer $i$ labels the $N$ different points on the contour. We use an algebraic method to obtain an accurate initial guess (Section \ref{Initial_guess}) for an iterative scheme to find the optimal circle and ellipse parameters (Section \ref{Optimal_parameters}).

\subsubsection{The initial guess for the model parameters}
\label{Initial_guess}

The estimated centre of the circle and ellipse are both at $\left( \overline{x}, \overline{y} \right)$, where
\begin{eqnarray}
    \overline{x} ~\equiv~ \frac{1}{N} \sum_i x_i \, , \quad \overline{y} ~\equiv~ \frac{1}{N} \sum_i y_i \, .
    \label{Mean_position_contour}
\end{eqnarray}
We subtract this from each point on the contour to obtain
\begin{eqnarray}
    \widetilde{x}_i ~\equiv~ x_i - \overline{x} \, , \quad \widetilde{y}_i ~\equiv~ y_i - \overline{y} \, .
\end{eqnarray}
These quantities are used to calculate the inertia tensor
\begin{eqnarray}
    && \matr{I} ~\equiv~ \begin{bmatrix} S_{xx} & S_{xy} \\
S_{xy} & S_{yy}
\end{bmatrix} \, , \quad \text{where} \\
    && S_{xx} ~\equiv~ \sum_i \widetilde{x}_i \widetilde{x}_i \, , ~S_{xy} ~\equiv~ \sum_i \widetilde{x}_i \widetilde{y}_i \, , ~ S_{yy} ~\equiv~ \sum_i \widetilde{y}_i \widetilde{y}_i \, . \nonumber
\end{eqnarray}
We then diagonalize $\matr{I}$ to find its eigenvalues $\lambda_1$ and $\lambda_2$, with $\lambda_1 > \lambda_2$. The aspect ratio of the ellipse is assumed to be
\begin{eqnarray}
    q ~=~ \sqrt{\frac{\lambda_2}{\lambda_1}} \, .
\end{eqnarray}
The semi-major axis of the ellipse is assumed to be along the eigenvector corresponding to $\lambda_1$. The square root is required here because the eigenvalues of the inertia tensor tell us the squared distance along any particular direction \citep[see also equation~11 of][]{Wu_2013}. There is also an extra factor of the total number of particles, but this cancels upon taking a ratio of eigenvalues.

Given the centre of the circular fit (Equation \ref{Mean_position_contour}), the only remaining parameter is its radius $r$. We estimate that
\begin{eqnarray}
    r ~=~ \sqrt{\frac{S_{xx} + S_{yy}}{N}} \, .
\end{eqnarray}
The semi-major axis $a$ of the ellipse is taken to be
\begin{eqnarray}
    a ~=~ \frac{r}{\sqrt{q}} \, .
\end{eqnarray}
This ensures that the initial guess circle and ellipse have the same area.

\subsubsection{Finding the optimal parameters}
\label{Optimal_parameters}

For each point on the contour, we find the minimum distance $d_i$ to the circle or ellipse by varying the azimuthal angle around it using a low-resolution grid search, which we refine using gradient descent \citep{Fletcher_1963}. Once this has converged, we quantify the goodness of the fit using
\begin{eqnarray}
    S ~\equiv~ \sum_i {d_i}^2 \, .
    \label{rms_error}
\end{eqnarray}
The root mean square (rms) error of the circle or ellipse fit is thus $\sqrt{S/N}$.

We optimize the ellipse fit using gradient descent in the five ellipse parameters (centre position, semi-major axis, aspect ratio, and orientation). The idea is the same for the best-fitting circle, though in this case we only need to consider three parameters (centre position and radius). For this work, the main parameters of interest are the aspect ratio $q_{\text{int}}$ of the best-fitting ellipse and the extent to which this is an improvement over the best-fitting circle.

\section{Results}
\label{Results}

In the early stages of their evolution, both the isolated and EFE models remain fairly circular. The models have a rotational period of 380~Myr at the mass-weighted average of $r_d$ and $r_g$, so we investigated snapshots after 3 rotational periods $-$ the non-axisymmetry mostly arises after this. Most snapshots in our analysis have $q_{_{\text{int}}} \leq 0.90$. To illustrate the possibility of a significantly elliptical contour, we only present results where $q_{_{\text{int}}} \leq 0.86$ for the contour at the level $\log_{10} \Sigma_{\text{gas}} = -0.6$ in units of $M_\odot/\textrm{pc}^2$, which we chose as representative of the disc outskirts. The contours at this level have a similar size to the isophote plotted in figure~7 of \citet{Mancera_2022}.

\begin{figure*}
    \includegraphics[width=0.495\textwidth]{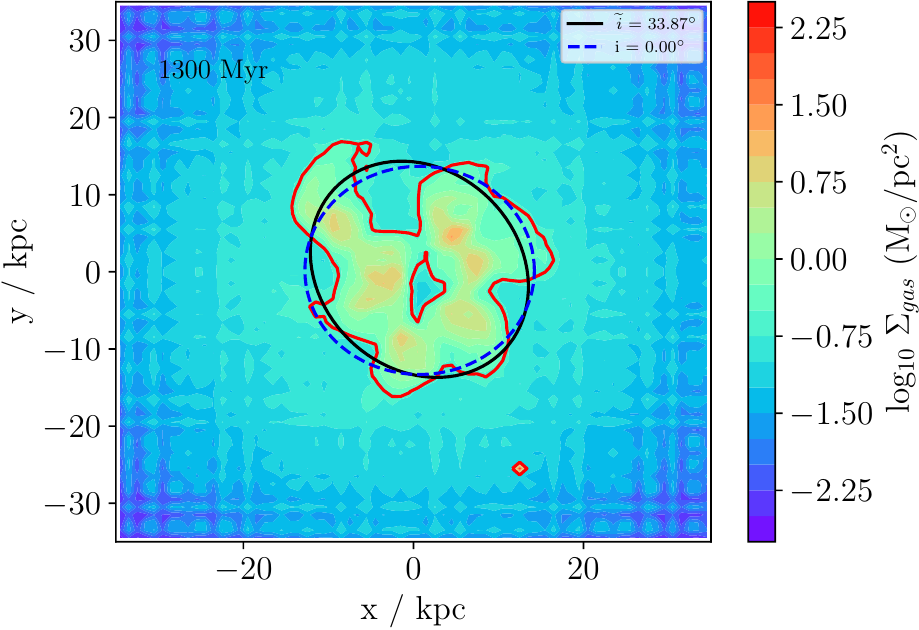}
    \hfill
    \includegraphics[width=0.495\textwidth]{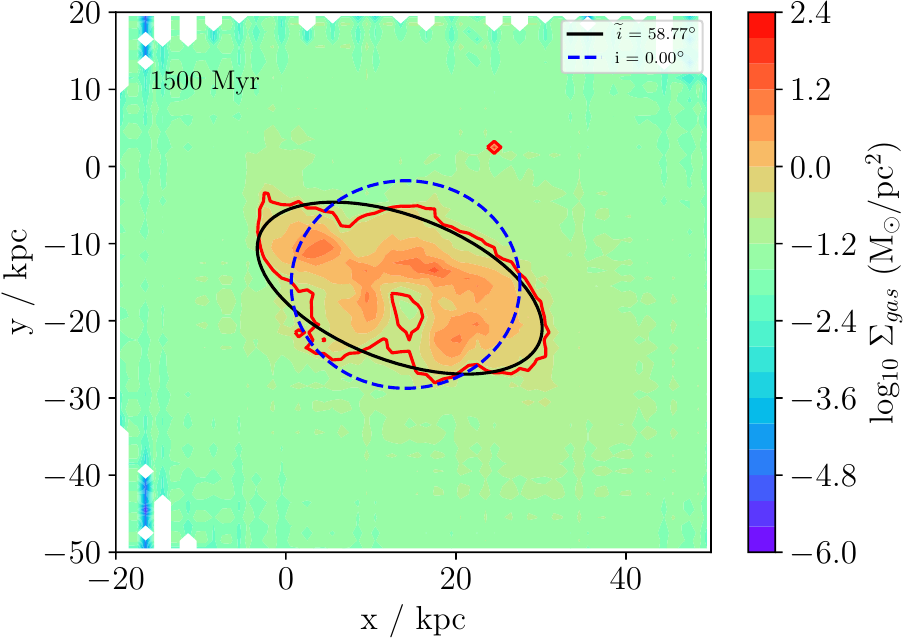}
    \includegraphics[width=0.495\textwidth]{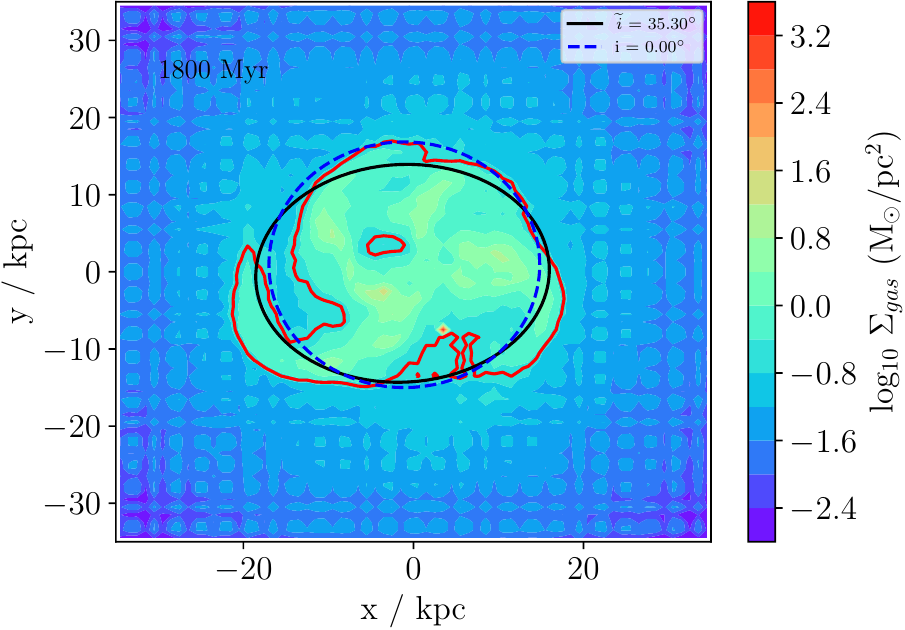}
    \hfill
    \includegraphics[width=0.495\textwidth]{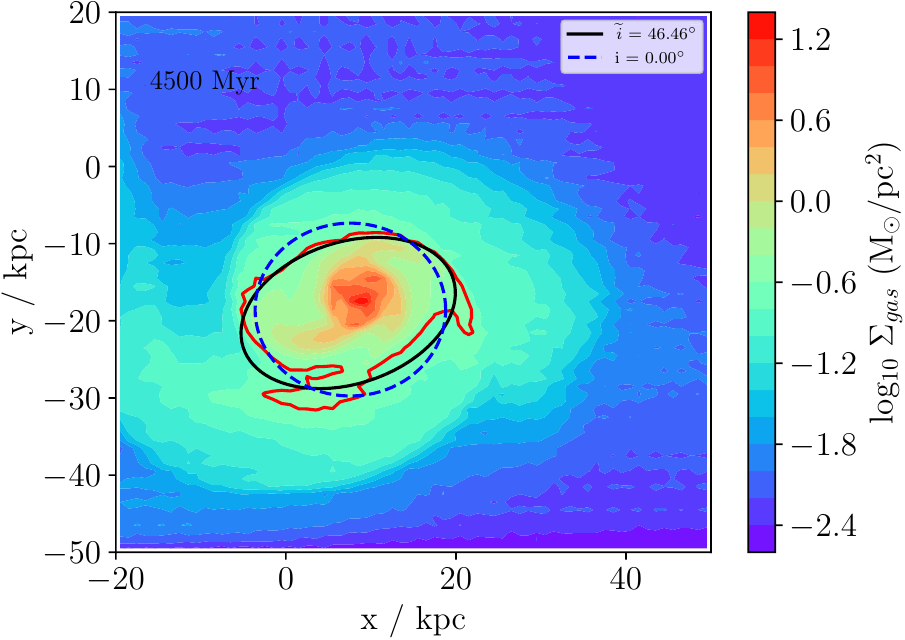}
    \includegraphics[width=0.495\textwidth]{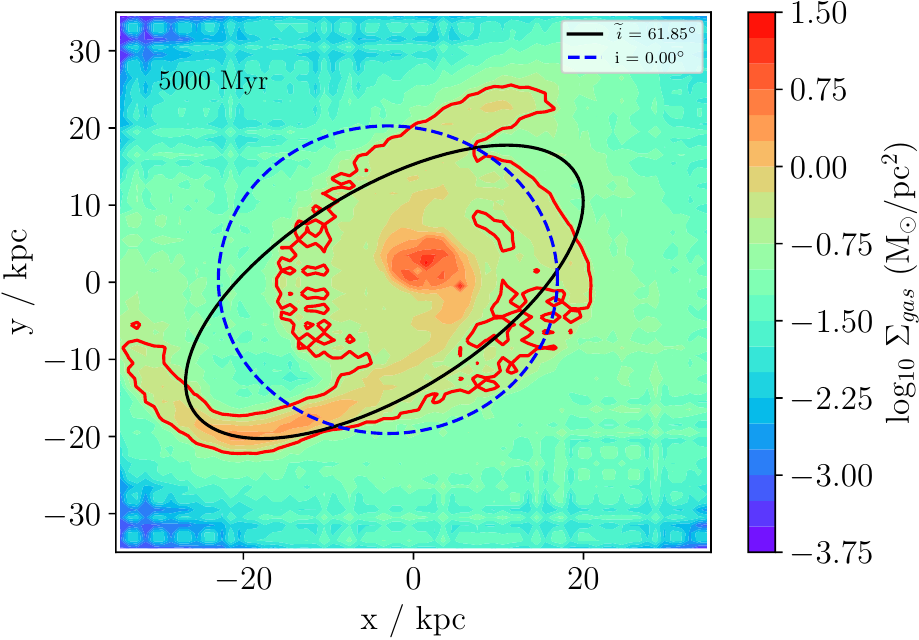}
    \hfill
    \includegraphics[width=0.495\textwidth]{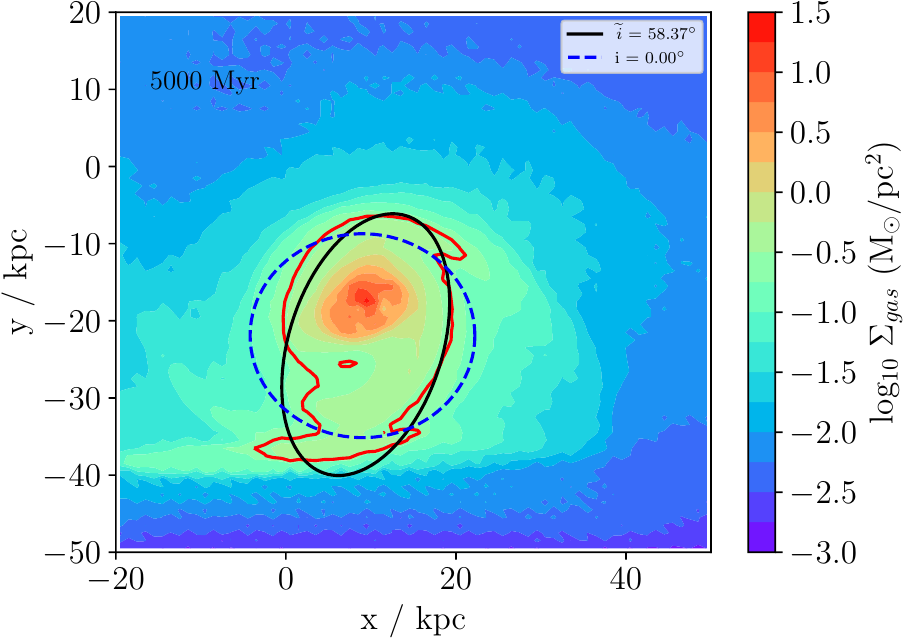}
    \caption{Galaxy models in isolation (\emph{left panels}) and including the EFE with $g_{\textrm{ext}} = 0.05 \, a_{_0}$ at $30^\circ$ to the disc plane but with no component along $y$ (\emph{right panels}), shown face-on ($i = 0^\circ$) at the indicated times. In the model with the EFE, the coordinate system has been rotated to view the galaxy along the direction of the total angular momentum because the EFE causes the disc to precess \citep[see section~4.2.5 of][]{Banik_2020_M33}. The colours show the gas surface density as $\log_{10} \Sigma_{\text{gas}}$, with $\Sigma_{\text{gas}}$ in units of $M_\odot$/pc\textsuperscript{2}. The solid red line shows the contour with level $-0.6$. The dotted blue curve shows the best-fitting circle. The aspect ratio $q_{\text{int}}$ of the best-fitting ellipse (solid black curve) is shown in the legend as $\widetilde{i} \equiv \cos^{-1} q_{\text{int}}$, the inclination that would be inferred by an observer seeing the galaxy face-on based only on $q_{\text{int}}$ (Equation~\ref{i_tilde}).}
    \label{Galaxy_snapshots}
\end{figure*}

In the isolated model, spiral structure starts developing after $\approx 2$~Gyr, with clear spiral features are present in all snapshots after 3~Gyr (see the left panels of Figure~\ref{Galaxy_snapshots}). Since AGC 114905 lacks clear spiral features, we present snapshots from this model at 1300~Myr and 1800~Myr. We also show the final snapshot at 5000~Myr, illustrating the spiral arms. These are generally two-armed, but some snapshots contain three-armed spirals, especially around the time when spiral structure is first apparent. This is in line with earlier MOND simulations \citep{Banik_2020_M33, Roshan_2021_disc_stability}.

\begin{figure*}
    \includegraphics[width=0.495\textwidth]{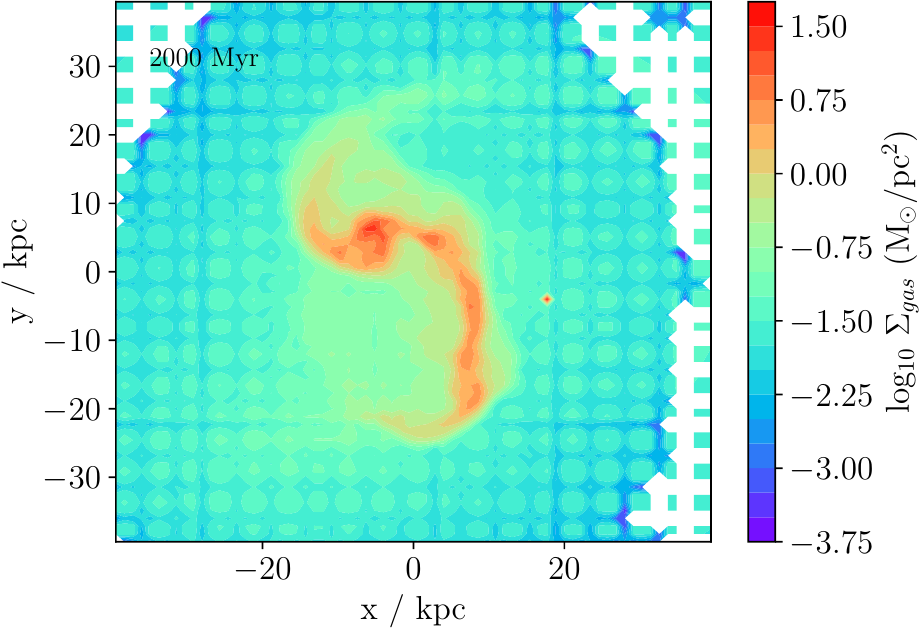}
    \hfill
    \includegraphics[width=0.495\textwidth]{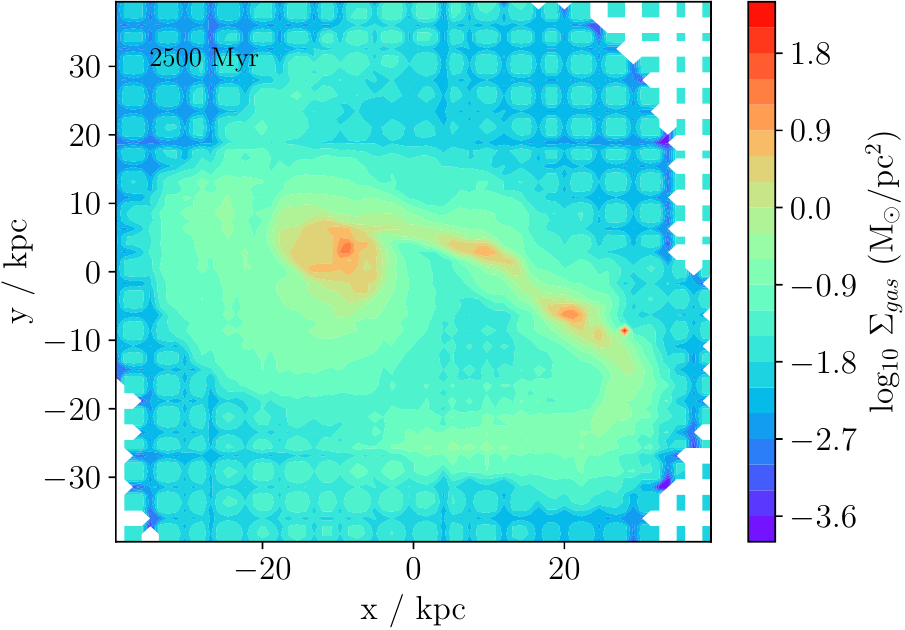}
    \includegraphics[width=0.495\textwidth]{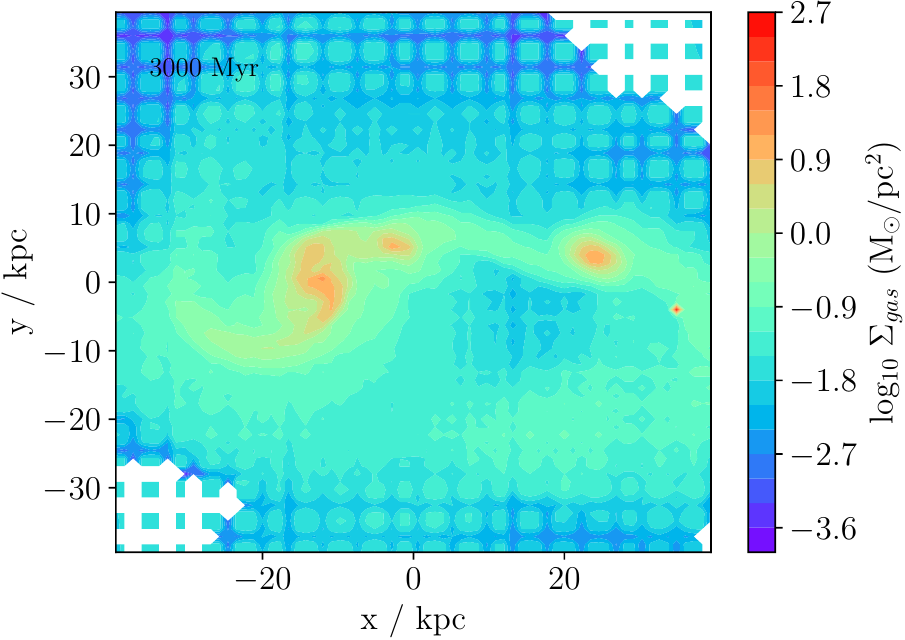}
    \hfill
    \includegraphics[width=0.495\textwidth]{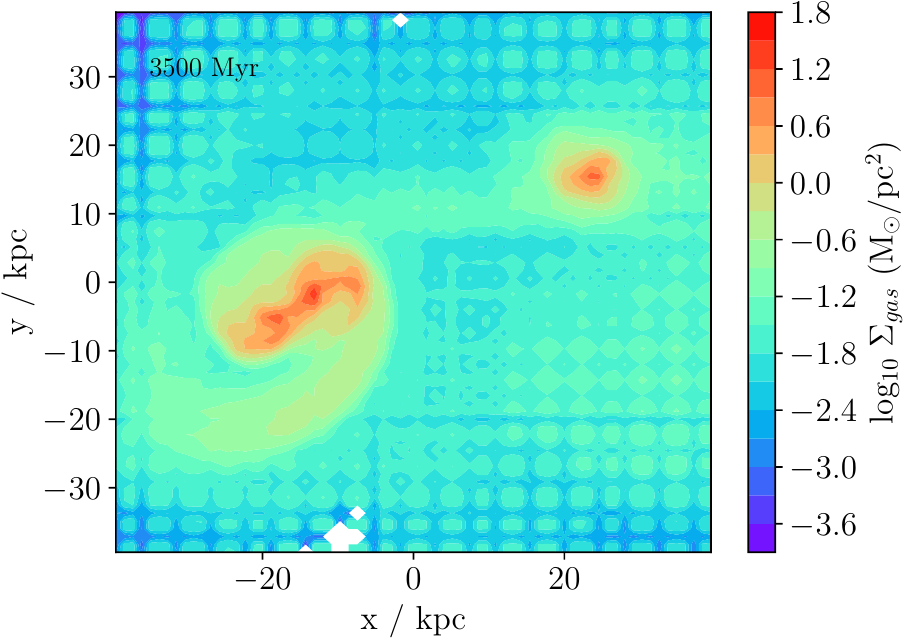}
    \caption{Snapshots of our galaxy simulation with the EFE, showing the formation of an overdense clump in the disc outskirts and its detachment to form a secondary galaxy.}
    \label{Galaxy_splitting}
\end{figure*}

After accounting for the much more substantial barycentre drift and disc precession (Section~\ref{Including_EFE}), the EFE model yields the snapshots shown in the right panels of Figure~\ref{Galaxy_snapshots}. A second clump of mass appears after 2000~Myr and then detaches from the galaxy by 3000~Myr $-$ the formation and detachment of this clump are illustrated in Figure~\ref{Galaxy_splitting}. However, the vast majority of the mass remains in the main clump: it has a mass of $0.99 \times 10^9\, M_{\odot}$ out of a combined mass in both clumps of $1.36 \times 10^9\, M_{\odot}$. The model looks fairly regular both in the 1500 Myr snapshot before the second clump is apparent and in the 4500~Myr snapshot after it has detached. We also show the final snapshot at 5000~Myr for completeness.

Despite viewing the galaxy face-on in both the isolated and EFE models, the best-fitting elliptical contour has an aspect ratio indicative of a disc inclined by $>32^\circ$, sometimes much more so (Table~\ref{Goodness_of_fit}). In the isolated model, the quality of the fit is however rather poor, as evidenced by the almost identical goodness of fit using the best-fitting circle and ellipse. In contrast, the disc in the EFE model has a much more regular appearance that is fit much better with an ellipse than with a circle. This is particularly so for the $\Sigma_{\text{gas}}$ contour in the 1500~Myr snapshot. Since the best-fitting ellipse has a significantly flattened shape, assuming the galaxy is axisymmetric would lead to a high inferred inclination of $59^\circ$. Moreover, there are no clear spiral arms that would alert observers to the possibility of a significantly overestimated inclination. This is also true with the 4500~Myr and 5000~Myr snapshots, though weak spiral arms are evident and the contour is not as well fit by an ellipse.

\begin{table}
    \begin{tabular}{ccccccc}
    \hline
    & & rms error & rms error & Aspect & \\ 
    & Snapshot & of circle & of ellipse & ratio \\
    Model & (Myr) & fit (kpc) & fit (kpc) & ($q_{\text{int}}$) & $\widetilde{i}$ \\ \hline
    Iso & 1300 & 3.10 & 3.15 & 0.87 & $33.9^\circ$ \\
    Iso & 1800 & 3.01 & 2.91 & 0.81 & $35.3^\circ$ \\
    Iso & 5000 & 6.49 & 4.27 & 0.47 & $61.9^\circ$ \\
    EFE & 1500 & 3.35 & 1.44 & 0.51 & $58.8^\circ$ \\
    EFE & 4500 & 2.04 & 1.57 & 0.68 & $46.5^\circ$ \\
    EFE & 5000 & 2.97 & 1.82 & 0.52 & $58.4^\circ$ \\ \hline
    \end{tabular}
    \caption{Parameters of the circle and ellipse fits to the contours shown on Figure~\ref{Galaxy_snapshots}. The columns show the simulation snapshot used, the rms error of the best-fitting circle and ellipse, and the aspect ratio of the latter. Assuming an intrinsically circular contour would lead to the inferred inclination being $\widetilde{i}$ (Equation~\ref{i_tilde}), even though the observer actually sees the galaxy face-on ($i = 0^\circ$).}
    \label{Goodness_of_fit}
\end{table}


The significant deviation from axisymmetry in our models is consistent with prior works: a similar deviation from axisymmetry can been in figure~8 of \citet{Banik_2020_M33} and figure~1 of \citet*{Wittenburg_2020}. Our work differs in that the central surface density of our model is much lower in order to match AGC 114905. However, the results remain qualitatively similar because even LSB discs are still completely self-gravitating in MOND, allowing them to sustain non-axisymmetric features like bars and spiral arms. Thus, relying merely on the shape of the isophote to estimate $i$ could well cause a significant overestimate.

\section{Discussion}
\label{Discussion}

The sky-projected aspect ratio $q_{\text{sky}}$ of an isophote can be used to estimate the inclination of a galaxy using
\begin{eqnarray}
    \cos \widetilde{i} ~\equiv~ q_{\text{sky}} \, ,
    \label{i_tilde}
\end{eqnarray}
where we use $\widetilde{i}$ to distinguish this estimate from the true $i$. The $\widetilde{i}$ values recovered in this way for face-on Milgromian disc galaxies are listed in Table~\ref{Goodness_of_fit}. For a galaxy observed at low inclination from slightly towards the minor axis, $q_{\text{sky}} = q_{\text{int}} \cos i$. Since $q_{\text{int}}$ can easily be smaller than $\cos 32^\circ/\cos 11^\circ = 0.86$, it is quite possible that a galaxy with $\widetilde{i} = 32^\circ$ actually has $i = 11^\circ$. Our results indicate that it could even be face-on.

This has significant implications for the RC estimated from line-of-sight velocities because deriving the RC within the disc plane requires a geometric correction factor of $1/\sin i$, which diverges at low $i$. The `observed' RC is based on a correction factor of $1/\sin \widetilde{i}$, which could be a lot smaller as discussed above. The actual RC could thus be equal to the RC reported by observers scaled up by a factor of $\sin \widetilde{i}/\sin i$, which in the above example would be 2.8. For the case of AGC 114905, this would raise the reported $v_{_f} = 23$ km/s up to 64 km/s, quite close to the MOND prediction that $v_{_f} = 69$ km/s (Equation~\ref{asymptoticvelocity}) for the estimated $M_b = 1.4 \times 10^9 M_\odot$ \citep{Mancera_2022}. The claimed inconsistency with MOND arises because those authors obtain $\widetilde{i} = 32^\circ$ from an elliptical isophote, as illustrated in their figure 7. Their argument is that the shape of the contour is such that $\widetilde{i}$ cannot be as low as $11^\circ$. However, the actual $i$ could certainly be this low in a Milgromian universe because the isophotes of a face-on galaxy may well be non-circular.

If as we argue AGC 114905 is nearly face-on, then even a small warp could significantly alter the shape of its inferred RC if this is not accounted for. Even rather isolated galaxies may become warped in MOND due to the EFE \citep{Brada_2000_warp, Banik_2020_M33}. Therefore, very nearly face-on galaxies should generally be excluded from RC studies. A cut of $i \geq 30^\circ$ was recommended in \citet{Kroupa_2018}, but their recommendation is for galaxies in the SPARC sample with much better observations that allow $i$ to be estimated using a tilted ring fit to the 2D velocity field \citep{Bosma_1978}. If only the shapes of the isophotes are used, our work suggests that a better criterion would be $i \geq 60^\circ$, which should provide some assurance that the observed ellipticity of the image is indeed due to inclination, especially if the isophotes can be fit well using ellipses with a fixed centre, axis ratio, and major axis direction.

\subsection{Overestimated inclinations in \texorpdfstring{$\Lambda$CDM}{LCDM}?}

At face value, the observed RC published by \citet{Mancera_2022} is also very problematic for $\Lambda$CDM, which like MOND prefers that $i = 11^\circ$ (see their section~5.4). It is therefore interesting to consider whether AGC 114905 could be nearly face-on in a $\Lambda$CDM context. This would require the apparent ellipticity to be a consequence of intrinsic non-axisymmetry in the face-on view, which can be addressed using numerical simulations. One problem is that if a galaxy is dominated by CDM, then its disc is essentially not self-gravitating, making it difficult to sustain bars and spiral arms. The fact that many LSB discs do have such features has been used to argue against $\Lambda$CDM \citep[see section 8 of][]{McGaugh_1998_LCDM}.

Nowadays, it is possible to consider dwarf galaxies in a large cosmological hydrodynamical simulation of the $\Lambda$CDM paradigm. Using zoom-in simulations, \citet{Marasco_2018} argued that triaxiality of the CDM halo could cause a disc to be somewhat non-circular when viewed face-on, even if the surface density is rather low. This is apparent in their figure~1, which considers two galaxies with identifier numbers 19 and 17. By considering the top row of this figure (galaxy 19), we found that the aspect ratio of the plotted best-fitting ellipse to an isophote of $\Sigma_{\star}$ is $\approx 0.95$, which would not be sufficient to explain AGC 114905 because $\cos 32^\circ/\cos 11^\circ = 0.86$. Moreover, we used the colours of the top left panel of this figure to estimate that the central $\Sigma_{\star} \approx 55 \, M_\odot/\textrm{pc}^2$, which is much higher than for AGC 114905 where $\Sigma_{\star} \approx 7 \, M_\odot/\textrm{pc}^2$ and $\Sigma_{\textrm{gas}} \approx 3 \, M_\odot/\textrm{pc}^2$ \citep[see figure~1 of][]{Mancera_2022}. We then considered the bottom row of figure~1 in \citet{Marasco_2018}, which shows their galaxy 17. In this case, the aspect ratio of the stellar component is $\approx 0.8$, which in principle would be sufficient to reconcile the apparent ellipticity of AGC 114905 with an almost face-on orientation. However, this galaxy has $\Sigma_{\star} \approx 80 \, M_\odot/\textrm{pc}^2$, which is almost an order of magnitude larger than the central baryonic surface density of AGC 114905. Since the above-mentioned simulated galaxies in the study of \citet{Marasco_2018} have a similar amount of mass in gas as in the stars (see their section~3.1), their baryonic surface density is much higher than for AGC 114905. Moreover, the dissipative gas component may well be even closer to axisymmetric than the stars. In summary, the results of \citet{Marasco_2018} apply to `typical' dwarf irregulars and thus cannot be extrapolated to the case of AGC 114905, which has a much lower stellar surface density and a larger disc than the simulated systems they studied. Further work is needed to clarify whether sufficiently non-circular gas discs with an extremely low surface density can spontaneously form in the $\Lambda$CDM framework.

Recently, this issue was investigated by \citet{Sellwood_2022} using Newtonian $N$-body simulations tailored to the properties of AGC 114905. In their model where $i = 15^\circ$ and the CDM halo is tuned to reproduce the resulting higher RC, the galaxy appears almost completely axisymmetric. However, $i = 15^\circ$ is already rather high: \citet{Mancera_2022} argue in their section~5.4 that AGC 114905 follows the expected relation between stellar and halo mass only if $i \approx 11^\circ$. Essentially, a low inclination is needed to reconcile the high expected RC amplitude with the small observed line-of-sight velocity gradient across AGC 114905. Using a lower inclination than in \citet{Sellwood_2022} would however make the disc even less relevant to the total potential, making it harder to understand how the isophotes can deviate from axisymmetry at the 14\% level. Thus, the preliminary results of \citet{Sellwood_2022} suggest that it is very difficult to generate a significant amount of non-axisymmetry in a galaxy like AGC 114905 where the CDM needs to dominate because $\Sigma_{\star} + \Sigma_{\textrm{gas}} \ll \Sigma_M$ (Equation~\ref{Sigma_M}). It is not yet clear whether considering a somewhat triaxial CDM halo could yield the required departure from axisymmetry if the degree of triaxiality and the orientation of the disc relative to the halo follow expectations in a broader cosmological context.

\section{Conclusions}
\label{Conclusions}

We conduct hydrodynamical MOND simulations of a gas-rich disc galaxy with a similar mass distribution to AGC 114905, first in isolation and then including the EFE with $g_{\textrm{ext}} = 0.05 \, a_{_0}$ directed at $30^\circ$ to the disc plane. We show that isodensity contours in $\Sigma_{\text{gas}}$ can appear to be elliptical at the 25\% level or more in the face-on view where $i = 0^\circ$. This could mislead observers into thinking that $i$ is as high as $60^\circ$. As RCs require a correction factor of $1/\sin i$, this could cause the true RC to greatly exceed that reported in the observational literature. In the specific case of AGC 114905, this invalidates the claim of \citet{Mancera_2022} that the low amplitude of its RC falsifies MOND because this claim relies on $i = 32^\circ$. The RC would be much higher and consistent with MOND if instead $i = 11^\circ$, which as shown in this work is quite possible. This is due to the fact that in MOND, even LSB discs are self-gravitating, allowing them to sustain non-axisymmetric features like bars and spiral arms \citep{Banik_2020_M33}. Observed LSB discs often do have such features \citep*{McGaugh_1995}. A related phenomenon in our simulations is that satellites can split off from the outskirts of a gas-rich disc galaxy in a process similar to the gravitational instability model of gas giant planet formation (Figure~\ref{Galaxy_splitting}). Such satellites should be orbiting close to the disc plane. It is intriguing that a galaxy not subject to tidal forces can nevertheless develop features resembling tidal tails.

Given the hundreds of RCs that agree with MOND \citep[e.g.,][]{Li_2018} and the inevitable outliers from the correct theory given imperfect data \citep[e.g.,][]{Kroupa_2018, Cameron_2020}, MOND is not presently in significant tension with galaxy RCs. We conclude that nearly face-on galaxies are unsuitable for such tests, especially in a theory where significant departures from axisymmetry are quite possible because galaxies are completely self-gravitating at any surface brightness.

\section*{Acknowledgements}

IB is supported by Science and Technology Facilities Council grant ST/V000861/1, which also partially supports HZ. The authors thank Ingo Thies, Moritz Haslbauer, Elena Asencio, and Jerry Sellwood for useful discussions. They are also grateful for comments from the anonymous referee which helped to substantially improve this publication.

\section*{Data availability}

The algorithms used to prepare and run \textsc{por} simulations of disc galaxies and to extract their results into human-readable form are publicly available.\footnote{\url{https://bitbucket.org/SrikanthTN/bonnpor/src/master/}} A user manual is available describing the operation of these codes \citep{Nagesh_2021}.

\bibliographystyle{mnras}
\bibliography{Fake_inclination_bbl}
\bsp
\label{lastpage}
\end{document}